\documentclass[aps,twocolumn,superscriptaddress,showpacs]{revtex4}
\usepackage{epsfig}
\usepackage{color}
\usepackage{graphicx}
\usepackage{dcolumn}
\usepackage{bm}
\usepackage[normalem]{ulem}
\usepackage{textcomp}
\usepackage{soul}

\mathchardef\mhyphen="2D

\begin{document}

\title{Correlated electronic structure, orbital-dependent correlations, and Lifshitz transition in tetragonal FeS}

\author{S. L. Skornyakov}

\affiliation{M. N. Miheev Institute of Metal Physics of Ural Branch of Russian Academy of Sciences, 18 S. Kovalevskaya Street, 620137 Yekaterinburg, Russia}
\affiliation{Ural Federal University, 620002 Yekaterinburg, Russia}

\author{I. Leonov}
\affiliation{M. N. Miheev Institute of Metal Physics of Ural Branch of Russian Academy of Sciences, 18 S. Kovalevskaya Street, 620137 Yekaterinburg, Russia}
\affiliation{Materials Modeling and Development Laboratory, National University of Science and Technology 'MISIS', 119049 Moscow, Russia}

\date{\today}

\begin{abstract}

Iron-based binary chalcogenide superconductors FeSe and FeS 
have attracted much recent attention due to their complex 
orbital-selective correlations and Cooper pairing, offering 
the minimal model system holding the key properties to 
understanding the physics of high-$T_c$ superconductors. 
Here, using density functional plus dynamical mean-field
theory method (DFT+DMFT) with full self-consistency over the 
charge density, we study the effect of electronic correlations 
on the electronic structure, magnetic properties, 
orbital-dependent band renormalizations, and Fermi surface 
of the tetragonal phase of bulk FeS. We perform a direct 
structural optimization of the $P_4/nmm$ crystal structure 
of paramagnetic FeS, minimizing the total energy of FeS with 
respect to the lattice constant $a$ and the internal coordinate 
$z_\mathrm{S}$ of atom S. Our results show an anomalous 
sensitivity of the electronic structure and magnetic properties 
of FeS to fine details of its crystals structure, e.g., to a 
small variation of the chalcogen coordinate $z_\mathrm{S}$. 
Upon expansion of the lattice volume (which can be realized, 
e.g., in FeS$_{1-x}$Se$_x$), we observe a remarkable change 
of the electronic structure of FeS which is associated with 
a complete reconstruction of the Fermi surface topology 
(Lifshitz transition). This behavior is ascribed to a 
correlation-induced shift of the Van Hove singularity 
associated with the Fe $xy$ and $xz/yz$ orbitals at the $M$ 
point across the Fermi level. The Lifshitz phase transition 
is accompanied by a significant growth of local magnetic 
moments and emergence of strong orbital-selective correlations. 
It is seen as a pronounced anomaly (`kink') in the total 
energies upon expansion of the lattice, associated with a 
remarkable enhancement of compressibility. This behavior is 
accompanied by an orbital-dependent formation of local moments, 
a crossover from itinerant to localized orbital-selective 
moment behavior of the Fe $3d$ electrons. While exhibiting 
weak effective mass enhancement of the Fe $3d$ states 
$m^*/m \sim 1.3-1.4$, correlation effects reveal a strong 
impact on a position of the Van Hove singularity originating 
from the Fe $xz/yz$ and $xy$ orbitals at the $M$ point, 
implying a complex interplay between electronic correlations 
and band structure effects in FeS. Our results suggest a 
complex interplay between electronic correlations, magnetism, 
and lattice degrees of freedom in FeS.

\end{abstract}

 \maketitle

\section{INTRODUCTION}

The discovery of unconventional superconductivity in the 
Fe-based pnictides and chalcogenides, with the highest 
transition temperature $T_c$ up to $55$~K in fluorine 
doped SmFeAsO and $\sim 109$~K in a monolayer of FeSe, 
has attracted a lot of attention from researchers around 
the world in the recent 
past~\cite{review_superconductors, 55K_Tc, FeSe_monolayer}. 
Similar to high-$T_c$ cuprates, manganites or some heavy 
fermion compounds, the Fe-based pnictide and chalcogenide 
superconductors (FeSCs) are characterized by the proximity 
and complex interplay of various phases, including magnetism, 
nematicity, and crystal structure phases~\cite{FeSe_rev_mag_nem}. 
In fact, superconductivity in FeSCs is often found to appear 
in the vicinity of a magnetic phase transition and/or 
structural (nematic) instability, as a result of the electron/hole 
doping, pressure or other means suppression of long-range, 
single-stripe antiferromagnetic (AFM) order with a wave 
vector $Q_{m}=(\pi,\pi)$~\cite{nematicity}. The Fermi surface 
topology of FeSCs is characterized by an in-plane nesting 
wave vector $(\pi,\pi)$, consistent with $s_\pm$ pairing 
symmetry~\cite{sign_reversal,pn_pairing}. Moreover, experimental 
studies of the spin excitation spectra of both pnictides 
and chalcogenides show an enhancement of short-range AFM 
spin fluctuations at a vector $(\pi,\pi)$ near the $T_c$~\cite{FeSe_pi_pi}. 
This behavior has been regarded as evidence for the importance 
of spin fluctuations in the pairing of electrons in FeSCs. 

Among various chemical compositions of FeSCs much 
attention has been paid to the binary chalcogenide 
compounds, the so-called '11' family of FeSCs, Fe\emph{X} 
with \emph{X}=Se, Te, and 
S~\cite{review_chalcogen_recent,FeSe_rev_spectra,FeSe_rev_mag_nem}. 
Besides their structural simplicity, the '11' FeSCs 
offer the minimal model system holding the key properties 
to understanding the physics of high-$T_c$ superconductors 
in general. Iron selenide FeSe is a particularly fascinating 
representative of the '11' FeSCs~\cite{review_chalcogen_recent, 
FeSe_rev_mag_nem, Sprau_Science_2017, Kostin_NatMat_2018}. 
Unlike the iron pnictides, FeSe lacks long range magnetic 
order in spite of a similar structural behavior and extended 
nematic (orthorhombic) phase below $90$~K~\cite{FeSe_rev_mag_nem,FeSe_ortho}. 
In fact, FeSe has been found to become superconducting below 
$T_c \sim 8$~K close to its stoichiometric composition~\cite{Superconductivity_FeSe}. 
Superconductivity in FeSe is highly tunable, with the 
superconducting transition temperature ranging 
from $T_c \sim 8$ K in bulk single crystals at ambient pressure to 
$\sim 40$~K under pressure (of $\sim 6$~GPa) or in intercalated 
systems, to $\sim 14$~K upon isoelectronic substitution 
of Se with Te (corresponding to a negative chemical pressure, 
i.e., lattice expansion) and to $\sim 65$--$109$~K in thin 
films~\cite{FeSe_hydrostatic, FeSe_hydrostatic_2, FeSe_Te_doping, FeSe_monolayer,FeSe_intercalation}.
While FeSe shows no antiferromagnetically ordered phase at 
ambient pressure down to the lowest temperatures, static 
magnetism (most likely) of the $(\pi,\pi)$ AFM stripe-type 
emerges upon compression above $\sim 1$~GPa~\cite{FeSe_pi_pi}. 
Moreover, a double-stripe $(\pi,0)$-type magnetically 
ordered phase appears upon expansion of the lattice caused 
by an isoelectronic substation of Se with Te, in FeTe~\cite{FeTe_pi_0}. 
The $(\pi,\pi)$ to $(\pi,0)$ crossover in magnetic correlations 
can be attributed to a Lifshitz transition of the electronic 
band structure of Fe(Se,Te), accompanied by a remarkable 
increase of local magnetic moments and an enhancement of 
electronic compressibility at the transition point~\cite{FeSeI,FeSeII,FeSeIII}. 
This behavior was ascribed to, e.g., a correlation-induced 
shift of the Van Hove singularity associated with the Fe 
$xy$ and $xz/yz$ orbitals at the Brillouin zone $M$ point 
across the Fermi level~\cite{FeSeI,FeSeII,FeSeIII}. 
Under pressure, $T_c$ in FeSe increases dramatically to a 
maximum of $\sim 37$~K at about $6$~GPa~\cite{FeSe_hydrostatic,FeSe_hydrostatic_2}. 
By contrast, substitution of Se for S, which constitutes a 
positive chemical pressure on FeSe, leads to a reduction of 
the degree of $(\pi,\pi)$ nesting, resulting in a decrease 
of $T_c$~\cite{FeS_nesting}. The latter is presumably connected 
with suppression of spin fluctuations, resulting in a lowering 
of the critical temperature in FeSe$_{1-x}$S$_x$ series~\cite{FeSeIII,FeSeS_ARPES}.

Being isoelectronic and isostructural counterpart of 
FeSe, the tetragonal FeS has attracted a lot of attention 
from both theory and experiment as a promising model 
system to study the underlying mechanisms (e.g., the 
interplay of magnetism and structural parameters) of 
superconductivity in FeSCs~\cite{FeS_allotropes}. In contrast to FeSe and the 
majority of FeSCs, FeS does not reveal a structural 
transition from the tetragonal to orthorombic structural 
phase (i.e., it exhibits no nematic order) down to the 
lowest temperatures~\cite{FeS_sc,FeSe_rev_mag_nem,FeSeS_ARPES}. 
Moreover, superconducting (SC) properties of FeS display 
a delicate interplay with the lattice. The $T_c$ is first 
found to decrease under pressure. However, upon further 
compression, it is followed by a reemergence of SC with 
formation of a second SC dome on the phase 
diagram~\cite{FeS_sc_suppress, FeS_two_domes}. This behavior 
can be attributed to a Lifshitz transition which adds 
a hole pocket to the Fermi surface (FS)~\cite{FeS_Lifshitz}. 
From angle-resolved photoemission experiments (ARPES) and 
electronic structure calculations the Fermi surface of FeS 
resembles that of FeSe~\cite{FeCh_DFT,FeS_ARPES,FeSeS_ARPES}. 
It features two electron-like pockets at the tetragonal 
Brillouin zone $M$ points and two hole-like pockets 
centered at the $\Gamma$ point, showing a moderate 
dispersion along $k_{z}$. However, at ambient pressure 
the third hole-like pocket at $\Gamma$ in FeS is not 
observed ~\cite{FeS_ARPES,FeSeS_ARPES}. In addition, 
ARPES shows that the increase of S content in the solid 
solution FeSe$_{1-x}$S$_x$ leads to an increase of the 
size of the FS, accompanied with a reduction of the degree 
of $(\pi,\pi)$ in-plane nesting~\cite{FeSeS_ARPES,FeS_nesting}. 
The latter suggests damping of spin fluctuations in FeS 
compared to those in FeSe. Photoemission and ARPES 
measurements of the electronic properties of FeS reveal 
a sufficient narrowing of the Fe $3d$ bandwidth as compared 
to band structure calculations~\cite{FeCh_DFT}. However, 
an orbital-dependent band renormalization of the quasiparticle 
mass is weaker than that in FeSe and is only about 
$\sim 1.5$--$2$, implying a remarkably weaker strength of 
correlation effects in FeS \cite{FeSeS_ARPES,FeS_dftdmft}. 
In this context it is interesting to note the possible 
importance of electron-phonon coupling in FeS~\cite{FeS_dft_phonon}. 

The most advanced theoretical methods for calculation of 
the electronic properties of strongly correlated materials, 
such as the density functional theory plus dynamical mean 
theory approach (DFT+DMFT) have shown to provide a good 
description of the electronic structure of 
FeSCs~\cite{dftdmft_nsc,dftdmft_sc,U_in_superconductors}. 
For FeSCs DFT+DMFT calculations reveal significant 
correlation effects with orbital-dependent strength and 
character, revealing the importance of the Hund's rule 
coupling induced many-body correlations~\cite{FeSeI,FeSeII, FeSeIII,U_in_superconductors,Skornyakov_2009,FeS_dftdmft,FeSeHubbard2,FeSe_compressibility}. 
This leads to substantial differences in quasiparticle 
weights and orbital selectivity of electronic correlations, 
suggesting that the Cooper pairing can also become 
orbital-selective, with a highly anisotropic superconducting 
gap \cite{Sprau_Science_2017,Kostin_NatMat_2018}. 
Moreover, such DFT+DMFT calculations indicate a positive 
correlation of the enhancement of the $T_c$ with the 
correlation-induced increase of the electronic compressibility, 
suggesting that the system is at the frontier of the normal-to-Hund's 
metal crossover~\cite{FeSeI,FeSeII,FeSeIII,FeSe_compressibility}. 
Applications of DFT+DMFT to the equilibrium phase of FeS and 
FeSe have shown that the electronic properties of FeS share 
many similarities with those of FeSe~\cite{FeS_dftdmft}. 
In agreement with ARPES measurements, the DFT+DMFT calculations 
suggest weaker orbital-dependent band renormalizations in 
FeS~\cite{FeSeS_ARPES,FeS_dftdmft}. In FeSe correlations 
are strong enough to induce a transfer of the spectral 
weight and form the lower Hubbard band whereas the existence 
of the low-energy correlation-induced satellite in tetragonal FeS 
remains an open question~\cite{FeSeHubbard1,FeSeHubbard2}.

As already discussed in many respects for FeSCs, the 
crystal structure effects (variation of the lattice 
under pressure and/or chemical substitutions) show a 
strong impact on the electronic properties and magnetic 
fluctuation spectrum of FeSCs, implying a complex 
coupling between the magnetic and structural properties~\cite{FeSe_rev_mag_nem}. 
It often results in a significant enhancement of the 
$T_c$'s upon a moderate variation of crystal structure 
parameters of FeSCs. This behavior still needs to be 
understood on a microscopic level that drives much 
attention to the physical properties of the '11' FeSCs. 
In this respect FeS, a superconducting system exhibiting 
no nematic ordering, offers new perspectives to study 
the coupling between the magnetic and structural properties 
in FeSCs. 

In this paper, we study the interplay of the electronic 
structure, magnetic properties, and the Fermi surface 
topology in the tetragonal (space group $P_4/nmm$) paramagnetic 
FeS upon variation of its crystal structure. In particular, 
by employing the DFT+DMFT method \cite{dftdmft_nsc,dftdmft_sc} 
we study the effect of the lattice structure, e.g., a 
change of the lattice volume and chalcogen coordinate $z_{\mathrm S}$ 
on the electronic properties, magnetic correlations, and the 
Fermi surface of FeS. Our results reveal the importance of 
electronic correlations for the electronic properties and 
structural optimization of FeS. For examples, we observe a 
strong impact of correlation effects on a position of the 
Van Hove singularity originating from the the Fe $xz$/$yz$ 
and $xy$ orbitals at the $M$ point. We show that the internal 
parameter $z_{\mathrm S}$ plays a key role for understanding 
the physical properties of FeS upon expansion of the lattice. 
Upon lattice expansion FeS is found to exhibit a structural 
collapsed-tetragonal to tetragonal phase transition, accompanied 
by a significant increase of local magnetic moments and an 
emergence of strong orbital-selective correlations near the 
transition point. The phase transition results in a remarkable 
increase of the electronic compressibility and is associated 
with a topological change of the Fermi surface (Lifshitz transition). 
The latter is accompanied by a crossover from itinerant to 
localized orbital-selective moment behavior. Our results clearly 
demonstrate the crucial importance of orbital-selective 
correlations for a realistic description of the electronic 
and lattice properties of FeS.

The paper is organized as follows. In Sec.~II we describe the 
technical details of our DFT+DMFT calculations. In Sec.~III~A 
we present results for  crystal structure optimization of FeS, 
focusing on the evolution of its electronic structure and local 
magnetic moments. Spectral properties and orbital-selective 
renormalizations of the Fe $3d$ states of FeS are discussed in 
Sec.~III~B. Our results for the evolution of the Fermi surface 
and spin susceptibility upon variation of the crystal structure 
parameters of FeS are shown in Sec.~III~C and D, respectively. 
Finally, the results are summarized in Sec.~IV.  

\section{METHOD}
Here, we study the effect of electronic correlation on the 
electronic structure, magnetic correlations, and crystal 
structure properties of tetragonal FeS (space group $P_4/nmm$) 
using the state-of-the-art fully self-consistent in charge 
density DFT+DMFT method~\cite{dftdmft_nsc,dftdmft_sc}. 
We perform a direct structural optimization of the tetragonal 
$P_4/nmm$ unit cell of FeS, minimizing the total energy of 
FeS with respect to the lattice constant $a$ and the internal 
coordinate of sulphur $z_{\mathrm S}$ ($z_{\mathrm S}$ is 
proportional to the height of S relative to the basal Fe 
plane). In these calculations, the $c/a$ ratio of the 
tetragonal unit cell was fixed to its experimental value 
$c/a = 1.367$ (at ambient pressure). To compute pressure, 
equilibrium lattice volume and bulk modulus we fit the 
calculated total energies using the third-order Birch-Murnaghan 
equation of state~\cite{Birch}. We note that the DFT+DMFT 
results exhibit a clear anomaly in the total energy behavior. 
Therefore we fit the equation of states separately for 
the low- and high-volume regions.

We have employed the DFT+DMFT approach implemented within 
the plane-wave pseudopotential formalism with generalized 
gradient approximation (GGA) for the DFT exchange-correlation 
potential~\cite{GGA}. The DMFT equations are solved in 
a basis set of the Wannier Fe $3d$ and O $2p$ orbitals. 
The latter are constructed using the atomic-centered 
symmetry-constrained Wannier functions defined over the 
full energy range spanned by the Fe $3d$ and the S $4p$ 
bands~\cite{WannierH}. We solve the realistic many-body 
problem using the continuous-time hybridization-expansion 
(segment) quantum Monte Carlo algorithm~\cite{ctqmc}. 
The calculations are performed in the paramagnetic state at 
an electronic temperature $T = 290$~K. The Coulomb interaction 
has been treated in the density-density approximation with the 
average Hubbard interaction $U = 3.5$~eV and Hund's exchange 
$J = 0.85$~eV for the Fe $3d$ shell, as estimated 
previously~\cite{U_in_superconductors}. The interaction 
parameters $U$ and $J$ are assumed to remain constant 
upon variations of the lattice. The spin-orbit coupling 
was neglected in our calculations. To account for the 
electronic interaction already described by DFT, we employ 
the fully localized double counting correction, evaluated 
from the self-consistently determined local occupations. 
Spectral properties and renormalizations of the effective 
electron mass of the Fe $3d$ orbitals are obtained from 
the real-axis self-energy $\hat\Sigma(\omega)$ computed 
using the Pad\'e analytical continuation procedure~\cite{Pade}. 

To quantify a degree of localization of the Fe $3d$ 
electrons of FeS upon variation of the lattice we compute 
the local spin-spin correlation function 
$\chi( \tau ) = \langle \hat{m}_z(\tau)\hat{m}_z(0) \rangle$ 
within DMFT, where $\hat{m}_z(\tau)$ is the instantaneous 
magnetization on the Fe $3d$ state at the imaginary time 
$\tau$. The evolution of magnetic correlations and possible 
magnetic instabilities of FeS upon variation of the lattice 
is analyzed by calculating the momentum-resolved static 
susceptibility in the particle-hole bubble approximation:

\begin{equation}
\label{eqn:chi}
\chi({\bf q})= -k_{\mathrm{B}}T\sum_{{\bf k},i\omega_{n}}\mathrm{Tr}
\hat G({\bf k},i\omega_{n}) \hat G({\bf k}+{\bf q},i\omega_{n}),
\end{equation}
where $\hat G({\bf k},i\omega_{n})$ is the interacting lattice 
Green’s function evaluated within DFT+DMFT, $T$ is the electronic 
temperature, and $\omega_n$ is the fermionic Matsubara frequencies. 

\section{RESULTS}


\subsection{Structural optimization and local magnetic moments}

\begin{figure}[t]
\centering
\includegraphics[width=0.65\textwidth,clip=true,angle=-90]{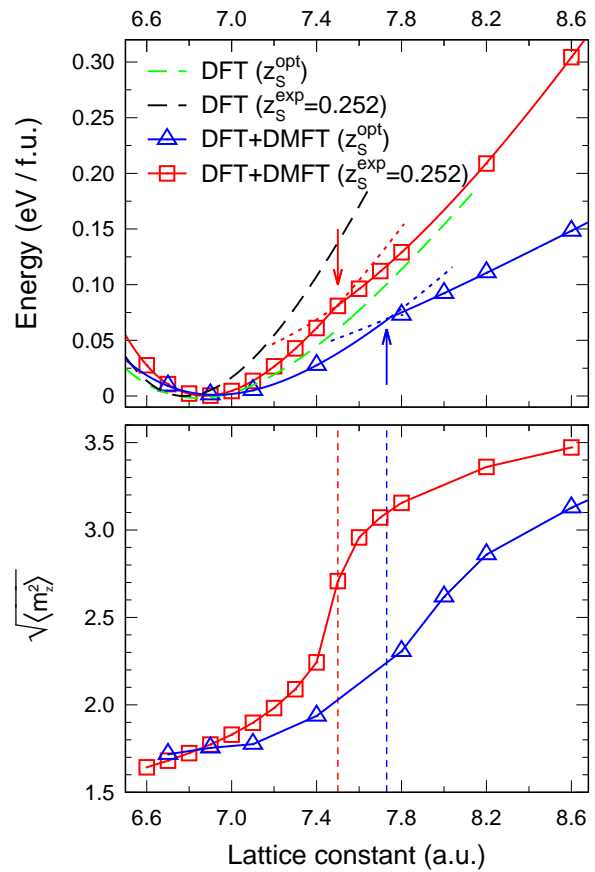}
\caption{
(Color online). Total energy (upper panel) and instantaneous 
local moment $\sqrt{\left<\hat{m}_z^2\right>}$ (lower panel) 
of paramagnetic FeS as a function of the lattice constant 
calculated by the charge self-consistent DFT+DMFT method at 
$T=290$~K for the experimental ($z_{\mathrm S}^{\mathrm{exp}}$) 
and optimized ($z_{\mathrm S}^{\mathrm{opt}}$) internal position 
of sulphur. The results for the total energy obtained within 
nonmagnetic GGA are shown in the upper panel for comparison 
(dashed curves). The phase transition point is indicated by 
arrows (upper panel) and by vertical dashed lines (lower 
panel).
}
\label{Fig_1}
\end{figure}

We start by performing a direct structural optimization 
of the $P_4/nmm$ crystal structure of paramagnetic FeS. 
To this end, we minimize the total energy of FeS with 
respect to the lattice constant $a$ and the internal 
coordinate $z_{\mathrm S}$ of atom S using the DFT+DMFT 
method with the average Coulomb interaction $U=3.5$~eV 
and Hund’s exchange $J=0.85$~eV for the Fe $3d$ shell. 
In Fig.~\ref{Fig_1} we compare our DFT+DMFT results 
with the results of the nonmagnetic GGA (nm-GGA) 
calculations (i.e., with the noninteracting case of 
$U=0$~eV and $J=0$~eV). Our nm-GGA results for the 
total energy show a typical parabolic-like behavior 
with the equilibrium lattice constant $a = 6.84$~a.u. 
and the fractional coordinate $z_{\mathrm S} = 0.241$ 
($a=6.78$~a.u. for the fixed $z_{\mathrm S}=0.252$). 
The calculated $a$ and $z_{\mathrm S}$ crystal structure 
parameters are by $\sim 2$\% and $4$\% smaller than 
those in the experiment, respectively. The calculated 
bulk modulus is found to depend very sensitively on the 
coordinate $z_{\mathrm S}$ (i.e., whether $z_{\mathrm S}$ 
is optimized or not). In particular, the bulk modulus 
computed from a fit of the DFT total energy to the 
third-order Birch-Murnaghan equation of state for the 
case with $z_{\mathrm S}$ fixed to its experimental 
value $z_{\mathrm S}=0.252$ is $K \sim 137$~GPa. 
The bulk modulus pressure derivative is 
$K^{\prime} \equiv dK/dP = 4.8$. On the other hand, 
structural optimization of FeS within nm-GGA with respect 
to both the lattice constant $a$ and coordinate $z_{\mathrm S}$ 
gives $K=67$~GPa. We note however that this value is 
still significantly higher (by about two times) 
than available experimental estimates for nanocrystalline 
tetragonal FeS ($\sim 30$~GPa)~\cite{FeS_bulkmod}. 
Interestingly, for the parent compound FeSe, which is 
isoelectronic and isostructural to FeS, experimental 
studies give $K \sim 31-33$~GPa \cite{FeSe_B0}. 

Structural optimization of the lattice parameters of 
paramagnetic FeS within DFT+DMFT gives an overall 
improvement for the calculated crystal structure 
parameters with respect to the non-interacting nm-GGA 
results. Our results for the lattice constant 
$a\sim 6.93$~a.u. (less than about 1\% off the experimental 
value) and the internal coordinate $z_{\mathrm S}=0.247$ 
(smaller by $\sim$2~\%) are in good agreement with 
available experiments. For the bulk modulus we obtain 
$K=106$~GPa (for $z_{\mathrm S}$ fixed to its experimental 
value), while simultaneous optimization of $a$ and 
$z_{\mathrm S}$ within DFT+DMFT gives a significantly 
softer crystal lattice with $K\sim 50$~GPa. The latter 
estimate is in reasonable agreement with the experimental 
bulk modulus of nanocrystalline tetragonal FeS 
($\sim 30$ GPa) and with the bulk modulus of FeSe 
($31$--$33$~GPa)~\cite{FeSe_B0,FeSe_structure}. We note 
that the calculated height of the chalcogen atom, sulphur 
in FeS, relative to the basal Fe plane of $2.332$~a.u. 
($z_{\mathrm S}=0.247$) is by $\sim15$\% smaller than that 
in the parent compound FeSe ($2.729$~a.u.), in agreement 
with the experiment. The calculated instantaneous local 
magnetic moment is 
$\sqrt{\langle \hat{m}_z^2 \rangle}= 1.75$ $\mu_{\mathrm B}$ 
(corresponding to the fluctuating local moment of 
$M_{loc}= 0.40$~$\mu_{\mathrm B}$, evaluated as 
$M_{loc} = (k_{\mathrm{B}}T\int_0^{1/k_{\mathrm{B}}T} d \tau \langle \hat{m}_z(\tau)\hat{m}_z(0)\rangle)^{1/2}$, 
where $T$ is the temperature) at ambient pressure. 
The calculated value of $\sqrt{\langle \hat{m}_z^2 \rangle}$ 
is by $\sim 8$~\% smaller as compared to that obtained 
for paramagnetic FeSe. Clearly, it is the inclusion of 
the local Coulomb interaction that provides an overall 
improved description of the properties of FeS compared 
to the nm-GGA results. 

Our results for the calculated crystal structure and 
electronic structure parameters of FeS are summarized in 
Table~\ref{Table1}. We find that the calculated crystal 
lattice parameters of FeS are in overall good agreement 
with recent experimental data. Most importantly, the 
DFT+DMFT calculations predict an isostructural phase 
transition which is accompanied by a substantial increase 
of the local magnetic moments $\sqrt{\left<\hat{m}_z^2\right>}$ 
(a formation of local moments) upon expansion of the 
lattice of FeS, i.e., under a "negative" (chemical) 
compression. Indeed, our calculations reveal a pronounced 
anomaly in the total energy upon expansion of the lattice, 
associated with a remarkable enhancement of electronic 
compressibility near $a \sim 7.5$--$7.7$~a.u. ($a \sim 7.5$~a.u. 
for fixed $z_{\mathrm S}$ and $\sim 7.7$~a.u. upon 
simultaneous optimization of $a$ and $z_{\mathrm S}$, 
respectively)~\cite{note}. Moreover, the instantaneous 
local moments which are  
$\sqrt{\left<\hat{m}_z^2\right>}\sim 1.75$~$\mu_\mathrm{B}$ 
at ambient conditions are found to increase to 
$3.36$~$\mu_{\mathrm B}$ and $2.86$~$\mu_{\mathrm B}$ at 
$a=8.2$~a.u., respectively (see lower panel of Fig.~\ref{Fig_1}). 
Interestingly, structural optimization of both $a$ and 
$z_{\mathrm S}$ results in a smooth evolution of the 
local moments as a function of volume, while the lattice 
anomaly shifts to a higher volume with $a \sim 7.8$~a.u.. 
We note that similar anomalous behavior of the lattice 
structure and local magnetic moments upon lattice expansion 
has been found in the parent system FeSe \cite{FeSeI,FeSeII,FeSeIII}. 
By analogy with FeSe, we interpret this behavior of FeS 
as a transition from a collapsed-tetragonal (equilibrium 
volume) to tetragonal (expanded volume) phase which occurs 
upon expansion of the lattice volume. The transition is 
of first order, occurs under a negative pressure of above 
$15$~GPa (in practice it can be realized, e.g., by substituting 
S with Se), which is as twice as that found in FeSe. 
The expanded-volume phase has a significantly smaller 
bulk modulus of $\sim 27$~GPa.

Our results show that the phase 
transition is accompanied by an increase of the lattice 
constant from $a = 7.3$~a.u. to $7.8$~a.u. (with fixed 
$z_{\mathrm S}=0.252$), as obtained from a common tangent 
construction for the energy-volume dependence. The latter 
corresponds to a large increase of the lattice volume by 
$\Delta V/V \sim 18$~\%. We note that optimization of 
$z_{\mathrm S}$ further lowers the energy of the expanded-volume 
phase and results in a significant extension of the 
transition region. This implies a significant variation 
of the $z_{\mathrm S}$ value at the transition. The phase 
transition spans the range from $a \sim 7.4$~ a.u. to 
about $\sim 8.6$~a.u., implying the importance of optimization 
of all crystal structure parameters, including $c/a$, for 
the expanded-volume phase of FeS. Moreover, similarly to 
FeSe this anomaly of the crystal structure of FeS is not 
observed in spin-polarized DFT calculations for the $(\pi,0)$ 
and $(\pi,\pi)$ antiferromagnetic ordering of Fe moments, 
demonstrating the importance of electronic correlations 
in FeS~\cite{FeSe_magnetic_DFT}.
\begin{table}[b]
\caption{
Calculated equilibrium structural parameters of FeS and 
the position of Van Hove singularity 
$E_{\mathrm{VHS}}^{\mathrm{DMFT}}$ at the $M$ point 
(relative to the Fermi energy) obtained by DFT+DMFT 
at $T=290$~K in comparison to the results for FeSe ~\cite{FeSeIII}. 
Here, ${h_{\mathrm{ch}}}$ is the height of a chalcogen 
atom ($\mathrm{ch}$=S, Se) above the basal Fe plane 
and $E_{\mathrm{VHS}}^{\mathrm{GGA}}$ is the position 
of VHS in nm-GGA.
}
\begin{tabular}{cccccc}
\hline
\hline
      &  $E_{\mathrm{VHS}}^{\mathrm{GGA}}$ & $E_{\mathrm{VHS}}^{\mathrm{DMFT}}$ & ${z_{\mathrm{ch}}}$ &  ${h_{\mathrm{ch}}}$  &  $c/a$   \\
      &  (eV)           & (eV)          &              &  (a.u.)       &            \\
\hline
 FeS  &  $-0.35$         & $-0.21$        & $0.247$      &  $2.332$      &  $1.367$   \\
 FeSe &  $-0.25$         & $-0.12$        & $0.261$      &  $2.729$      &  $1.494$   \\
\hline
\hline
\end{tabular}
\label{Table1}
\end{table}


\subsection{Spectral properties and band renormalizations}

\begin{figure}[t]
\centering
\includegraphics[width=0.37\textwidth,clip=true,angle=-90]{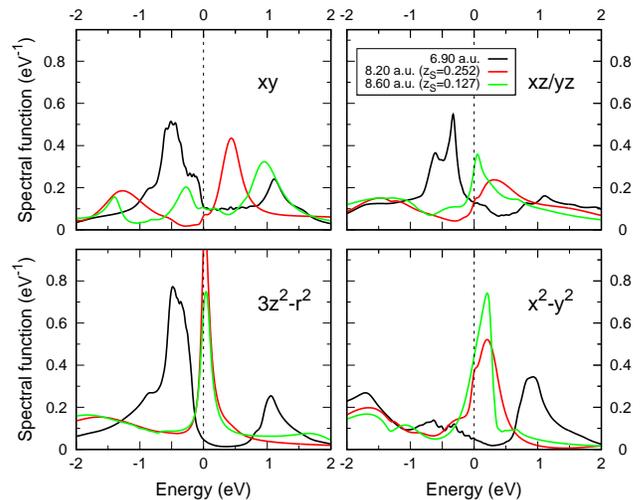}
\caption{
(Color online). Orbitally-resolved Fe $3d$ spectral functions 
of FeS computed by DFT+DMFT at $T=290$~K for the equilibrium 
lattice volume (black) and those in the expanded-volume phase 
with experimental $z_{\mathrm S}=0.252$ (red) and optimized 
$z_{\mathrm S}=0.127$ (green). The Fermi energy is shown by 
a vertical dashed line ($E_{\mathrm F}=0$~eV).
}
\label{Fig_2}
\end{figure}

\begin{figure}[t]
\centering
\includegraphics[width=0.44\textwidth,clip=true,angle=-90]{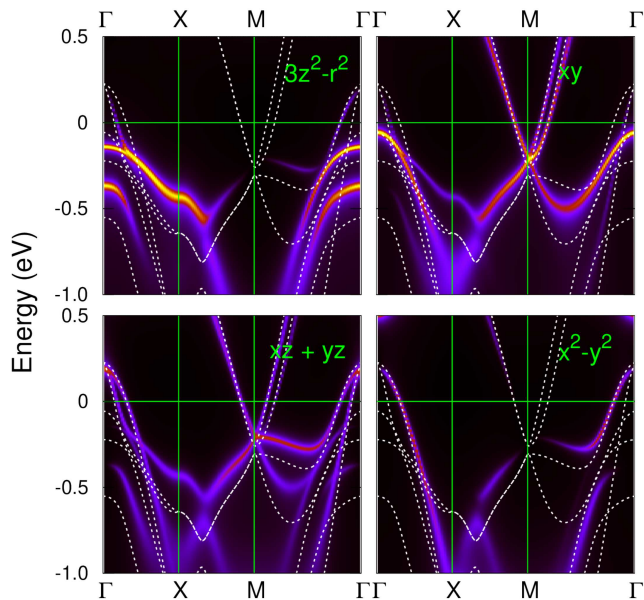}
\caption{
(Color online). Band structure of FeS along the $\Gamma\mhyphen X\mhyphen M\mhyphen \Gamma$ 
path obtained by DFT+DMFT at ambient pressure ($a=6.90$~a.u., 
$z_{\mathrm S}=0.252$) for $T=290$~K. The dashed curves show 
the nm-GGA results.
}
\label{Fig_3}
\end{figure}

Next, we discuss the spectral properties of paramagnetic 
FeS. In Fig.~\ref{Fig_2} we show the spectral functions 
of FeS calculated for the equilibrium structure and for 
the high-volume phase ($a=8.20$~a.u. for the fixed 
$z_{\mathrm{S}}=0.252$ and $a=8.60$~a.u. for the optimized  
$z_{\mathrm{S}}=0.127$). Our results for the electronic 
band structure of FeS obtained within DFT+DMFT for the 
calculated equilibrium structural parameters $a=6.90$~a.u. 
and $z_{\mathrm{S}}=0.252$ for are shown in Fig.~\ref{Fig_3}. 
We notice that the spectral functions of FeS are 
qualitatively similar to those calculated earlier for 
the parent compound FeSe. Moreover, in qualitative 
agreement with FeSe, the Fe $xy$ and $xz/yz$ spectral 
functions exhibit a well-defined quasiparticle peak (QP)
located below the Fermi level at about $-0.32$~eV. Our 
analysis of the correlated band structure of FeS suggests 
that this peak is originating from the Van Hove singularity 
(VHS) of the $xy$ and $xz/yz$ bands located below $E_{\mathrm{F}}$ 
at about $-0.21$~eV at the $M$ point. In addition, the 
spectral functions for the $t_2$ orbitals of FeS exhibit 
a broad feature at $\sim-1.5$~eV, associated with the 
lower Hubbard band. For comparison, the DFT+DMFT calculations 
for FeSe place the QP associated with the Fe $xy$ and $xz/yz$ 
VHS at $\sim-0.2$~eV below the Fermi level. 

We find a substantial renormalization of the Fe $3d$ 
bands with respect to the nm-GGA results, which is 
however sufficiently weaker than in FeSe. In fact, our 
results for the orbital-selective renormalization of 
the Fe $3d$ states evaluated from 
$m^{*}/m=1-\partial\Sigma(\omega)/\partial\omega|_{\omega\to 0}$ 
(here, $\Sigma(\omega)$ is the self-energy in the 
real-frequency domain computed by employing Pad\'e 
approximants \cite{Pade}) exhibit a rather weak 
enhancement of $m^{*}/m$, ranging in $1.3$--$1.4$ 
(see Table \ref{Table2}). This implies that at ambient 
conditions FeS is a weakly correlated system with small 
orbital-selective renormalizations. The Fe $t_2$ states 
are being renormalized by about $1$\% stronger than the 
$e$ states. We note that the DFT+DMFT results for FeSe 
in the vicinity of the equilibrium lattice volume give 
$m^{*}/m$ lying in the range $1.5$--$2$. In accordance 
with this, our DFT+DMFT calculations show that the Fe $3d$ 
bandwidth of FeS is by $\sim 10$\% larger as compared to 
that of FeSe, implying overall weaker correlation effects 
in FeS. Nonetheless, we point out that the effect of 
correlations on the electronic structure of FeS is 
non-negligible, resulting in a significant shift of the 
quasiparticle bands in the vicinity of the Fermi level. 
In particular, we observe a remarkable shift of the Fe 
$xz$/$yz$ and $xy$ VHS at the $M$ point towards the 
Fermi level by about $40$\%, caused by correlation 
effects. The VHS shifts from $-0.35$~eV in the nm-GGA 
to $-0.21$~eV below the Fermi energy in the DFT+DMFT 
calculation (see Fig.~\ref{Fig_3}). Interestingly, in 
FeSe the correlation-induced shift of VHS towards the 
Fermi energy is even stronger. In particular, the VHS 
shifts from $-0.24$~eV (in nm-GGA) to $-0.12$~eV in the 
equilibrium phase of FeSe. 

\begin{figure}[t]
\centering
\includegraphics[width=0.35\textwidth,clip=true,angle=-90]{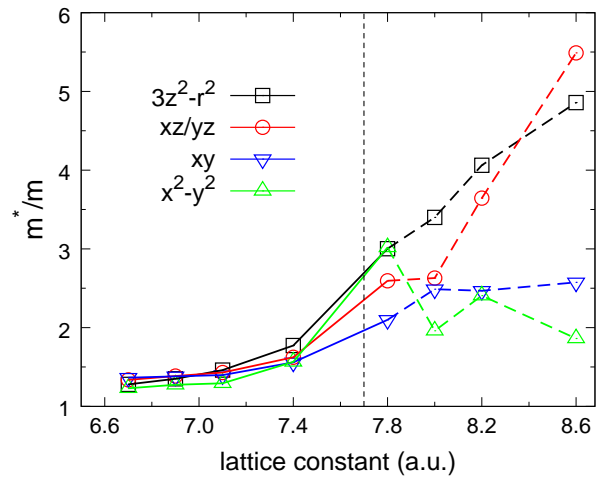}
\caption{
(Color online). Orbitally-resolved quasiparticle mass 
enhancement $m^{*}/m$ of the Fe $3d$ states in paramagnetic 
FeS as a function of lattice constant calculated by DFT+DMFT 
with optimized $z_{\mathrm S}$ at $T = 290$~K. The phase 
transition point is indicated by a vertical dashed line.
}
\label{Fig_4}
\end{figure}

In addition, we notice a 
remarkable correlation between the position of VHS and 
the strength of electronic correlation effects in FeSe 
and FeS (see Fig.~\ref{Fig_4} for FeS and Fig.~7 in 
Ref.~\cite{FeSeII} for FeSe). In fact, our results show 
that both FeS and FeS exhibit a significant orbital-dependent 
increase of $m^*/m$ to $\sim 3$ upon the VHS approaching 
the Fermi level (upon expansion of the lattice). While 
considering a `blue-red shift' problem in FeSCs (an 
inappropriate shifting upward of the electron bands at 
the $M$ point and downward of the holelike bands at the 
$\Gamma$ point, resulting in a shift of VHS away from 
$E_{\mathrm F}$) this rises important question about the 
interplay of VHS and correlation effects in FeSCs~\cite{Borisenko_VHS}.

Upon expansion of the lattice, we find a substantial 
redistribution of the Fe $3d$ spectral weight near the 
Fermi energy. We note that the overall change of the spectral 
function shape upon expansion of the lattice agrees with 
the evolution of photoemission spectra of Fe(Se,Te) obtained 
upon substitution of Se with Te \cite{FeSeTe_PES}. We observe 
that a substantial part of the spectral weight of the Fe 
$xz/yz$ and $xy$ orbitals in the vicinity of $E_{\mathrm{F}}$ 
is pushed from below to above the Fermi level. In particular, 
the QP peak associated with the VHS at $-0.23$~eV below the 
Fermi level in the equilibrium volume phase is absent for 
larger volumes. The $x^2-y^2$ spectral weight shifts in the 
opposite direction, towards $E_{\mathrm{F}}$, from $\sim 1$~eV 
to $0.25$~eV above $E_{\mathrm{F}}$. The $3z^2-r^2$ states 
form a quasiparticle peak near the Fermi level in the 
high-volume phase. 

The spectral weight transfer is accompanied by a remarkable 
orbital-selective renormalization of the Fe $3d$ bands. 
In fact, the effective mass of the Fe $t_2$ states increases 
to $\sim 5.49$ for the $xz/yz$ and $2.58$ for the $xy$ orbitals, 
i.e., almost by about $4$ times and by $2$ times with respect 
to the equilibrium values (see Table~\ref{Table2}). Moreover, 
we find that the effective mass of the Fe $3d$ states depends 
very sensitively on the lattice constant $a$ and coordinate 
$z_{\mathrm{S}}$. For example, our results show a significant 
change of $m^{*}/m$ for the $xz/yz$ and $xy$ orbitals, from 
$m^{*}/m=7.04$ and $6.94$ for $z_{\mathrm S}=0.127$ to 3.64 
and 2.47 for $z_{\mathrm S}=0.157$, respectively. In addition, 
we find a significant renormalization of the $3z^2-r^2$ orbitals, 
from $m^{*}/m \sim 1.35$ for the equilibrium lattice to $4.86$ 
for the high-volume phase with $a=8.60$~a.u. and $z_{\mathrm{S}}=0.127$. 
In Table~\ref{Table2} we also present our results for the 
orbital-dependent occupations of the Fe $3d$ states. Interestingly, 
the calculated orbital-selective renormalization of the Fe $3d$ 
states is accompanied by a change of the corresponding Fe $3d$ 
occupancies towards the half-filled state.

\begin{table}[b]
\caption{
Orbitally-resolved enhancement of the effective mass $m^{*}/m$
(top) and occupations of the Fe $3d$ states per spin (bottom)
of FeS for different lattice parameters as computed by the
charge self-consistent DFT+DMFT method at $T=290$~K.
}
\begin{tabular}{cccccc}
\hline
\hline
\multicolumn{2}{c}{Structural parameters} & \multicolumn{4}{c}{$m^*/m$} \\
  $a$ (a.u.)           &   $z_{\mathrm S}$& ${3z^2-r^2}$ &  ${xz/yz}$  &  ${xy}$   &  ${x^2-y^2}$ \\
\hline
$6.90$ & $0.252$ &  $1.36$        &  $1.41$       &  $1.42$     &  $1.29$        \\
$8.20$ & $0.252$ &  $3.65$        &  $7.04$       &  $6.94$     &  $2.50$        \\
$8.20$ & $0.157$ &  $4.06$        &  $3.64$       &  $2.47$     &  $2.41$        \\
$8.60$ & $0.127$ &  $4.86$        &  $5.49$       &  $2.58$     &  $1.86$        \\
       &         & \multicolumn{4}{c}{occupancies (per spin)}\\
$6.90$ & $0.252$ &  0.81   &  0.69   &  0.72   & 0.73    \\
$8.20$ & $0.252$ &  0.74   &  0.61   &  0.60   & 0.76    \\
$8.20$ & $0.157$ &  0.69   &  0.69   &  0.63   & 0.73    \\
$8.60$ & $0.127$ &  0.70   &  0.67   &  0.62   & 0.71    \\
\hline
\hline
\end{tabular}
\label{Table2}
\end{table}


\subsection{Fermi surface and Lifshitz transition}

\begin{figure}[t]
\centering
\includegraphics[width=0.45\textwidth,clip=true,angle=0]{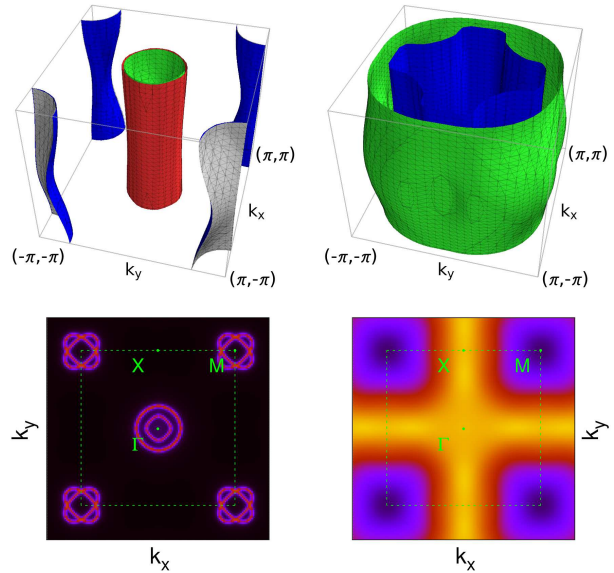}
\caption{
Fermi surface (upper panels) and spectral weight at the Fermi 
level (lower panels) of tetragonal FeS as computed by DFT+DMFT 
for $a=6.90$~a.u. (left) and $a=8.20$~a.u.~(right) with fixed 
z$_{\mathrm{S}}=0.252$.
}
\label{Fig_5}
\end{figure}

\begin{figure}[!h]
\centering
\includegraphics[width=0.45\textwidth,clip=true,angle=0]{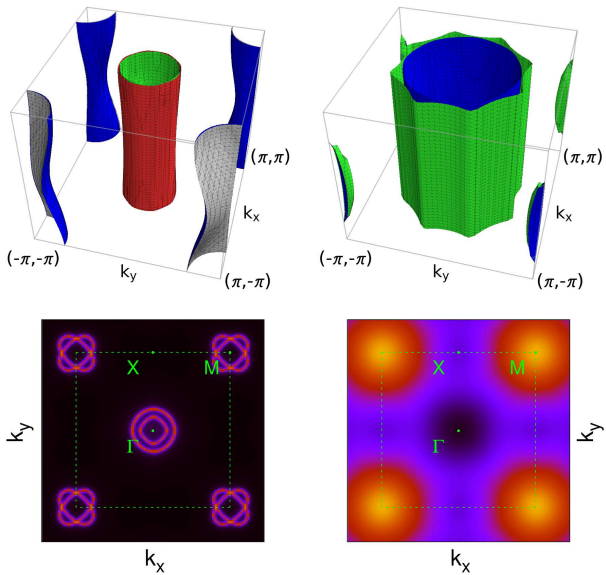}
\caption{
Fermi surface (upper panels) and spectral weight at the Fermi 
level (lower panels) of tetragonal FeS as computed by DFT+DMFT 
for $a=6.90$~a.u. (left) and $a=8.60$~a.u. (right) with optimized 
z$_{\mathrm{S}}$ (z$_{\mathrm{S}}$=0.247 and 0.127, respectively).
}
\label{Fig_6}
\end{figure}

To understand the effects of lattice expansion and 
electronic correlations in more detail, we analyze the 
$\bf k$-resolved spectral properties of FeS. For this 
purpose, we determine the poles of the DFT+DMFT lattice 
Green's function $\hat G({\bf k},\omega)$ (evaluated 
using $\hat\Sigma(\omega)$) and compute the momentum-resolved 
spectral function
$A({\bf k},\omega)=-\frac{1}{\pi}{\mathrm{Tr}}\hat G({\bf k},\omega)$. 
The former allows us to visualize the quasiparticle Fermi 
surface. On the other hand, the spectral function $A({\bf k},\omega)$ 
can be used to analyze the spectral weight at $E_{\mathrm F}$. 
In Figs.~\ref{Fig_5} and \ref{Fig_6} we show our results 
for the spectral weight at the Fermi energy (top row  
in Figs.~\ref{Fig_5} and \ref{Fig_6}) and the quasiparticle 
Fermi surface (bottom) obtained by DFT+DMFT for the different 
structural parameters $a$ and $z_{\mathrm S}$. For the 
equilibrium lattice parameters, the Fermi surface exhibits 
two hole sheets centered at the $\Gamma\mhyphen Z$ direction 
and two FSs at each of the four corners ($M\mhyphen Z$ 
direction) of the tetragonal Brillouin zone (BZ). 
The cross-section area of the corner FSs shows a strong 
dependence on $k_{z}$, reaching its minimum for $k_{z}=0$ 
and a maximum at the BZ boundary. The Fermi surface of 
the outer hole cylinder (mostly originating from a combination 
of the Fe $xz/yz$ and $x^2-y^2$ states) is almost parallel 
to the $\Gamma\mhyphen Z$ direction, i.e., two-dimensional. 
By contrast, the inner FS pocket has a less pronounced 2D 
character with a sufficient variation of its cross-section 
along $k_z$. Our results for the spectral weight show coherent 
FSs, implying a small damping of quasiparticles in the 
equilibrium phase. This is consistent with a weak renormalization 
of the Fe $3d$ states in the equilibrium phase.

Upon expansion of the lattice, we observe an entire 
reconstruction of the Fermi surface topology (Lifshitz transition) 
of FeS. With fixed $z_{\mathrm S}=0.252$, the FSs at the 
corners of the BZ vanish, while those centered at $\Gamma\mhyphen Z$ 
change their shape and significantly increase in size. 
In particular, now there are two hole-like quasi-2D FSs 
at $\Gamma\mhyphen Z$. The shape of the outer sheet is 
cylinder-like at the zone boundary $k_{z}=\pi$, with 
insufficient expansion along the $(\pi,\pi,k_{z})$ diagonals 
near $k_{z}=0$ (see Fig.~\ref{Fig_5}). The inner FS is 
almost parallel to $\Gamma\mhyphen Z$ and shows a cross-like 
cross-section at $k_{z}=0$. We note that in the 
expanded-volume phase the spectral weight distribution 
at the $E_{\mathrm F}$ is highly incoherent. This is mostly 
due to the inner FS pocket and is seen as four spots at 
$(\pi,0)$. The incoherent spectral weight at $E_{\mathrm F}$ 
implies a stronger damping of quasiparticles in the high-volume 
phase, in agreement with a larger band renormalization $m^{*}/m$. 
In this case, the overall evolution of the FS shape resembles 
that obtained within DFT+DMFT for FeSe \cite{FeSeI,FeSeII}.

We also note that our results for the FS of the high-volume 
phase of FeS depend sensitively upon the lattice coordinate 
$z_{\mathrm S}$. In particular, for the optimized value 
$z_{\mathrm S}=0.127$ the electron-like pockets centered 
at the BZ corners become three-dimensional like, closing 
at a halfway from $k_{z}=0$ to $k_{z}=\pi$. The cross-section 
area of the hole FS at the BZ center becomes significantly 
larger as compared to that in the equilibrium phase, weakly 
depending on $k_{z}$. In addition, we observe that the outer 
sheet shows concave structures at its surface, whereas the 
inner sheet has a cylindrical shape (see Fig.~\ref{Fig_6}). 
The spectral weight at $E_{\mathrm F}$ is highly incoherent 
and mostly originates from the electron-like FSs at the BZ 
corners. Interestingly, whether or not the $z_{\mathrm S}$ 
is optimized, the lattice volume expansion results in a 
change of the FS topology, i.e., the Lifshitz transition 
in FeS. We therefore conclude that upon expansion of the 
lattice ("negative compression") the Lifshitz transition 
takes place in tetragonal FeS. The phase transition is 
accompanied by anomalous behavior of the crystal structure 
and local magnetic moments, resulting in a remarkable 
enhancement of electronic compressibility. Our results are 
therefore in line with the scenario of a correlation-induced 
shift of the VHS originating from the Fe $t_2$ orbitals at 
the $M$ point towards the Fermi energy in FeSe \cite{FeSeI, FeSeII, FeSeIII}. 
The latter seems to lead to an anomalous behavior of the 
electronic structure, magnetism, and lattice properties of 
FeS upon variation of its lattice.


\subsection{Spin susceptibility and orbital-selective local moments}
We now turn to the magnetic properties of FeS and discuss 
our results for the spin susceptibility and orbital-selective 
local moments in paramagnetic FeS. In Fig.~\ref{Fig_6} we show 
our results for the orbital-dependent spin susceptibility
$\chi(\tau)=\langle \hat{m}_z(\tau)\hat{m}_z(0)\rangle$ computed 
within DFT+DMFT. We found that the electronic and structural phase 
transition upon expansion of the lattice volume is accompanied 
by a significant growth of the fluctuating local magnetic moments. 
The transition results in a crossover from itinerant 
(at ambient conditions) to orbital-selective localized moment 
behavior (in the high-volume phase), as is seen from the spin 
correlation function. In fact, our calculations reveal that the 
expanded-volume phase exhibits an orbital-selective enhancement 
of localization of the Fe $3d$ electrons. Interestingly, we found 
that the symmetry of magnetic correlations depends sensitively 
on a change of the lattice parameter $z_{\mathrm S}$ value, 
whereas the overall tendency to form local moments remains 
the same. In particular, upon simultaneous optimization of $a$ 
and $z_{\mathrm S}$, we obtain that the $3z^2-r^2$ and $xz/yz$ 
states in FeS show more localized behavior (to form fluctuating 
local moments) than that for the $x^2-y^2$ and $xy$ orbitals. 

\begin{figure}[t]
\centering
\includegraphics[width=0.60\textwidth,clip=true,angle=-90]{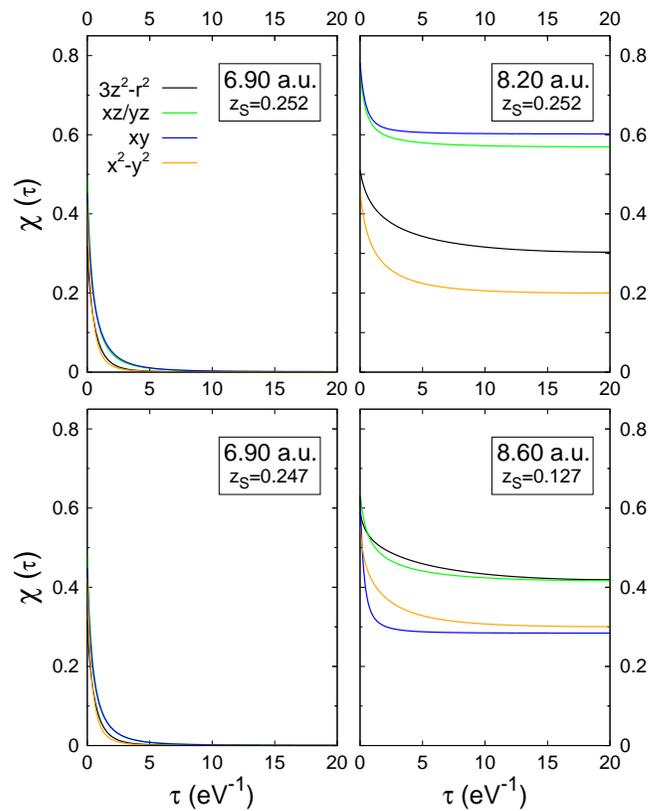}
\caption{
Orbitally-resolved local spin correlation functions 
$\chi(\tau)=\left< m_{z}(\tau)m_{z}(0)\right>$ of FeS 
at the equilibrium (left panels) and those in the expanded 
phase (right panels) as computed by DFT+DMFT at $T=290$~K. 
Top row shows the result obtained with fixed internal 
coordinate $z_{\mathrm S}$ bottom row corresponds 
to $z_{\mathrm S}$ optimized within DFT+DMFT.}
\label{Fig_7}
\end{figure}

\begin{figure}[t]
\centering
\includegraphics[width=0.55\textwidth,clip=true,angle=-90]{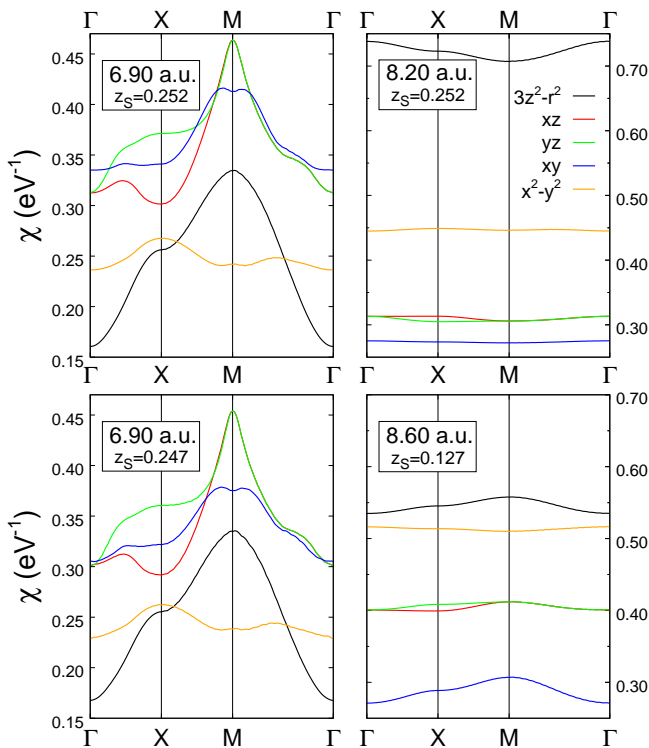}
\caption{
Orbitally-resolved static spin susceptibility $\chi({\bf q})$ of FeSe 
calculated by DFT+DMFT at fixed (top panels) and relaxed (bottom panels) 
internal coordinate $z_{\mathrm S}$. Left panels correspond to the 
equilibrium lattice constant, right panels show the DFT+DMFT results 
for the expanded phase.
}
\label{Fig_8}
\end{figure}

To analyze the effect of lattice expansion on the symmetry and 
strength of magnetic fluctuations we compute the momentum-dependent 
static magnetic susceptibility $\chi({\bf q})$. Orbital contributions 
of $\chi({\bf q})$ along the $\Gamma\mhyphen X\mhyphen M\mhyphen \Gamma$ 
path are shown in Fig.~\ref{Fig_8}. At ambient conditions, the 
susceptibility resembles that computed earlier for the equilibrium 
phase of FeSe. In particular, $\chi({\bf q})$ exhibits a maximum 
at the $M$ point of the tetragonal Brillouin zone, implying that 
leading magnetic instability at ambient pressure in FeS is due to 
the $t_2$ states, with a propagating wave vector $(\pi,\pi)$. 
However, we found that unlike to FeSe the largest contribution at 
the $M$ point is now originating from the $xz/yz$ orbitals but not 
from the $xy$ states as in FeSe. We also note that in the equilibrium 
phase of FeS the amplitude of spin fluctuations quantified by 
$\chi({\bf q})$ is overall smaller than that in FeSe. Overall, 
these results suggest that there exists a crossover of magnetic 
correlations in Fe(Se,S) upon substituting Se with S.

Our results show that expansion of the lattice volume is 
accompanied by a reconstruction of magnetic correlations.
Thus, upon expansion of the lattice we observe an overall 
enhancement of $\chi({\bf q})$. In the expanded-volume phase 
$\chi({\bf q})$ has a less pronounced variance along the 
$\Gamma\mhyphen X\mhyphen M\mhyphen \Gamma$ path than that 
in the equilibrium phase. Moreover, the Fe $e$ states now 
play a predominant role with a leading contribution originating 
from the $3z^2-r^2$ orbitals. Upon expansion of the lattice, 
the magnetic susceptibility shows a significant damping of 
a peak of $\chi({\bf q})$ at the $M$ point. As a result, 
$\chi({\bf q})$ reveals a weak feature at the $\Gamma$ point, 
implying a possible importance of ferromagnetic fluctuations 
(for $z_{\mathrm S}=0.252$). While $z_{\mathrm S}$ is optimized, 
a leading magnetic instability has a magnetic vector $(\pi,\pi)$, 
i.e., similar to that in the equilibrium phase of FeS, with 
a major contribution from the $xy$ and $3z^2-r^2$ states. 
Overall, our results point out an anomalous sensitivity of 
the electronic structure and magnetic properties of FeS to 
the fine details of its crystals structure, e.g., to a variation 
of the lattice volume and lattice coordinate $z_{\mathrm S}$. 
The latter results in a remarkable orbital-selective localization 
of the Fe $3d$ electrons.

\section{Conclusion}
In conclusion, using the DFT+DMFT method we calculated the 
evolution of the electronic structure, magnetic properties, 
and the Fermi surface of FeS upon variation of its crystal 
structure parameters. We explore the effects of the crystal 
structure -- changes of the lattice volume and chalcogen height 
$z_{\mathrm S}$ -- on the electronic properties, orbital-dependent 
character and strength of magnetic correlations, and the Fermi 
surface topology of the tetragonal (space group $P_4/nmm$) FeS. 
Our results show a clear importance of electronic correlations 
for the electronic properties and structural optimization of FeS. 
Correlation effects reveal a strong impact on a position of the 
Van Hove singularity originating from the the Fe $xz$/$yz$ and 
$xy$ orbitals at the $M$ point. In particular, we found that 
in the equilibrium structure, the QP peak associated with VHS 
is shifted towards the Fermi level from $-0.35$ to $-0.23$~eV. 
Our results suggest that the internal coordinate of sulfur, 
$z_{\mathrm S}$, plays a key role for understanding the physical 
properties of FeS upon variation of the lattice volume. Upon lattice 
expansion, FeS is found to exhibit a structural collapsed-tetragonal 
to tetragonal phase transition, accompanied by a significant 
increase of local moments and emergence of strong orbital-selective 
correlations near the transition point. The phase transition is 
associated with a topological change of the Fermi surface 
(Lifshitz transition), resulting in a remarkable increase of 
the electronic compressibility at the transition. Our DFT+DMFT 
calculations exhibit an anomalous sensitivity of the electronic 
structure and magnetic properties of FeS to the fine details 
of its crystals structure, e.g., to a variation of the lattice 
volume and lattice coordinate $z_{\mathrm S}$. Our results clearly 
demonstrate the crucial importance of orbital-selective correlations 
for a realistic description of the electronic and lattice properties 
of FeS, highlighting the importance of a crossover from itinerant 
to localized orbital-selective moment behavior of the Fe $3d$ electrons.


\section{ACKNOWLEDGMENTS}

Total energy calculations and analysis of magnetic correlations 
were performed with the support from the Russian Foundation for 
Basic Research according to the research project \textnumero~18-32-20076. 
The spectral functions, Fermi surface and the spin correlator 
calculations were supported by the Russian Science Foundation 
(Project \textnumero~19-12-00012).



\begin{thebibliography}{99}

\bibitem{review_superconductors}
J. Paglione and R. L. Greene,
Nat. Phys. {\bf 6}, 645 (2010);
%
D. N. Basov and A. V. Chubukov,
Nat. Phys. {\bf 7}, 272 (2011);
%
G. R. Stewart,
Rev. Mod. Phys. {\bf 83}, 1589 (2011);
%
P. Dai, J. Hu, and E. Dagotto,
Nat. Phys. {\bf 8}, 709 (2012);
%
Q. Si, R. Yu, and E. Abrahams,
Nat. Rev. Mater. \textbf{1}, 16017 (2016).


\bibitem{55K_Tc}
Z.-A. Ren, W. Lu, J. Yang, W. Yi, X.-L. Shen, Zheng-Cai, G.-C. Che, 
X.-L. Dong, L.-L. Sun, F. Zhou, and Z.-X. Zhao, 
Chin. Phys. Lett. {\bf 25}, 2215 (2008).


\bibitem{FeSe_monolayer}
S. Tan, Y. Zhang, M. Xia, Z. Ye, F. Chen, X. Xie, R. Peng, D.
Xu, Q. Fan, H. Xu, J. Jiang, T. Zhang, X. Lai, T. Xiang, J. Hu,
B. Xie, and D. Feng, Nat. Mater. {\bf 12}, 634 (2013).


\bibitem{FeSe_rev_mag_nem}
A. E. B\"ohmer and A. Kreisel, 
J. Phys.: Condens. Matter {\bf 30}, 023001 (2018).


\bibitem{nematicity}
R. M. Fernandes, A. V. Chubukov, and J. Schmalian, 
Nat.Phys. {\bf 10}, 97 (2014).


\bibitem{sign_reversal}
I. I. Mazin, D. J. Singh, M. D. Johannes, and M. H. Du, 
Phys. Rev.Lett. {\bf 101}, 057003 (2008).


\bibitem{pn_pairing}
A. V. Chubukov, D. V. Efremov, and I. Eremin, 
Phys. Rev. B {\bf 78}, 134512 (2008).


\bibitem{FeSe_pi_pi}
A. D. Christianson, E. A. Goremychkin, R. Osborn, S. Rosenkranz, M. D. Lumsden, C. D. Malliakas, I. S. Todorov, H. Claus, D. Y. Chung, M. G. Kanatzidis, R. I. Bewley, and T. Guidi;
Nature (London) {\bf 456}, 930 (2008);
%
M. D. Lumsden, A. D. Christianson, D. Parshall, M. B. Stone,
S. E. Nagler, G. J. MacDougall, H. A. Mook, K. Lokshin,
T. Egami, D. L. Abernathy, E. A. Goremychkin, R. Osborn,
M. A. McGuire, A. S. Sefat, R. Jin, B. C. Sales, and D. Mandrus,
Phys. Rev. Lett. {\bf 102}, 107005 (2009);
%
Y. Qiu, W. Bao, Y. Zhao, C. Broholm, V. Stanev, Z. Tesanovic,
Y. C. Gasparovic, S. Chang, Jin Hu, Bin Qian, Minghu Fang,
and Zhiqiang Mao,
Phys. Rev. Lett. {\bf 103}, 067008 (2009);
%
M. D. Lumsden, A. D. Christianson, E. A. Goremychkin,
S. E. Nagler, H. A. Mook, M. B. Stone, D. L. Abernathy,
T. Guidi, G. J. MacDougall, C. de la Cruz, A. S. Sefat,
M. A. McGuire, B. C. Sales, and D. Mandrus,
Nat. Phys. {\bf 6}, 182 (2010);
%
M. C. Rahn, R. A. Ewings, S. J. Sedlmaier, S. J. Clarke, and A. T. Boothroyd, 
Phys. Rev. B {\bf 91}, 180501(R) (2015).


\bibitem{review_chalcogen_recent}
X. Liu, L. Zhao, S. He, J. He, D. Liu, D. Mou, B. Shen,
Y. Hu, J. Huang, and X. J. Zhou,
J. Phys.: Condens. Matter {\bf 27}, 183201 (2015);
%
Y. V. Pustovit and A. A. Kordyuk,
Low Temp. Phys. {\bf 42}, 995 (2016);


\bibitem{FeSe_rev_spectra}
A. I. Coldea and M. D. Watson, 
Annu. Rev. Condens. Matter Phys. {\bf 9}, 125 (2018); 


\bibitem{Kostin_NatMat_2018}
A. Kostin, P. O. Sprau, A. Kreisel, Y. X. Chong, A. E. B\"ohmer, 
 P. C. Canfield, P. J. Hirschfeld, B. M. Andersen, and J. C. S\'eamus Davis, 
 Nat. Mater. \textbf{17}, 869 (2018).


\bibitem{Sprau_Science_2017}
P. O. Sprau, A. Kostin, A. Kreisel, A. E. B\"ohmer, V. Taufour, 
P. C. Canfield, S. Mukherjee, P. J. Hirschfeld, B. M. Andersen, 
J. C. S\'eamus Davis, Science \textbf{357}, 75 (2017).


\bibitem{FeSe_ortho}
T.M. McQueen, A. J. Williams, P. W. Stephens, J. Tao, Y. Zhu,
V. Ksenofontov, F. Casper, C. Felser and R. J. Cava, 
Phys. Rev. Lett. {\bf 103}, 057002 (2009).


\bibitem{Superconductivity_FeSe}
F. C. Hsu, J. Y. Luo, K. W. Yeh, T. K. Chen, T. W. Huang, P. M. Wu,
Y. C. Lee, Y. L. Huang, Y. Yi. Chu, D. C. Yan and M. K. Wu,
Proc. Natl. Acad. Sci. U.S.A. {\bf 105}, 14262 (2008).


\bibitem{FeSe_Te_doping}
B. C. Sales, A. S. Sefat, M. A. McGuire, R. Y. Jin,
D. Mandrus, and Y. Mozharivskyj,
Phys. Rev. B {\bf 79}, 094521 (2009);
%
V. Tsurkan, J. Deisenhofer, A. G{\"u}nther, Ch. Kant, H.-A. Krug von Nidda,
F. Schrettle, A. Loidl,
Eur. Phys. J. B \textbf{79}, 289-299 (2011);
%
U. R. Singh, S. C. White, S. Schmaus, V. Tsurkan, A. Loidl,
J. Deisenhofer, and P. Wahl,
Phys. Rev. B \textbf{88}, 155124 (2013).


\bibitem{FeSe_hydrostatic}
K. Miyoshi, K. Morishita, E. Mutou, M. Kondo, O. Seida,
K. Fujiwara, J. Takeuchi, and S. Nishigori,
J. Phys. Soc. Jpn. {\bf 83}, 013702 (2014).


\bibitem{FeSe_hydrostatic_2}
S. Medvedev, T. M. McQueen, I. A. Troyan, T. Palasyuk, M. I.
Eremets, R. J. Cava, S. Naghavi, F. Casper, V. Ksenofontov, G.
Wortmann, and C. Felser, Nat. Mater. {\bf 8}, 630 (2009).


\bibitem{FeSe_intercalation}
M. Burrard-Lucas, D. G. Free, S. J. Sedlmaier, J. D. Wright,
S. J. Cassidy, Y. Hara, A. J. Corkett, T. Lancaster, P. J. Baker,
S. J. Blundell, and S. J. Clarke, Nat. Mater. {\bf 12}, 15 (2013).


\bibitem{FeTe_pi_0}
W. Bao, Y. Qiu, Q. Huang, M. A. Green, P. Zajdel,
M. R. Fitzsimmons, M. Zhernenkov, S. Chang, M. Fang,
B. Qian, E. K. Vehstedt, J. Yang, H. M. Pham,
L. Spinu, and Z. Q. Mao,
Phys. Rev. Lett. {\bf 102}, 247001 (2009);
%
T. J. Liu, J. Hu, B. Qian, D. Fobes, Z. Q. Mao,
W. Bao, M. Reehuis, S. A. J. Kimber, K. Proke\v s,
S. Matas, D. N. Argyriou, A. Hiess, A. Rotaru,
H. Pham, L. Spinu, Y. Qiu, V. Thampy, A. T. Savici,
J. A. Rodriguez, and C. Broholm,
Nat. Mater. {\bf 9}, 718 (2010);
%
O. J. Lipscombe, G. F. Chen, C. Fang, T. G. Perring,
D. L. Abernathy, A. D. Christianson, T. Egami, N. Wang,
J. Hu, and P. Dai,
Phys. Rev. Lett. {\bf 106}, 057004 (2011).


\bibitem{FeSeI}
I. Leonov, S. L. Skornyakov, V. I. Anisimov, and D. Vollhardt,
Phys. Rev. Lett. {\bf 115}, 106402 (2015);


\bibitem{FeSeII}
S. L. Skornyakov, V. I. Anisimov, D. Vollhardt, and I. Leonov,
Phys. Rev. B {\bf 96}, 035137 (2017).


\bibitem{FeSeIII}
S. L. Skornyakov, V. I. Anisimov, D. Vollhardt, and I. Leonov,
Phys. Rev. B {\bf 97}, 115165 (2018).


\bibitem{FeS_nesting}
M. D. Watson, T. K. Kim, A. A. Haghighirad, S. F. Blake,
N. R. Davies, M. Hoesch, T. Wolf, and A. I. Coldea,
Phys. Rev. B {\bf 92}, 121108(R) (2015).


\bibitem{FeSeS_ARPES}
P. Reiss, M. D. Watson, T. K. Kim, A. A. Haghighirad, 
D. N. Woodruff, M. Bruma, S. J. Clarke, and A. I. Coldea,
Phys. Rev. B \textbf{96}, 121103(R) (2017);
%
A. I. Coldea, S. F. Blake, S. Kasahara, A. A. Haghighirad,
M. D. Watson, W. Knafo, E. S. Choi, A. McCollam,
P. Reiss, T. Yamashita, M. Bruma, S. C. Speller,
Y. Matsuda, T. Wolf, T. Shibauchi, and A. J. Schofield,
npj Quantum Materials {\bf 4}, 2 (2019).






\bibitem{FeS_allotropes}
Interestingly, FeS is also known for its rich allotropic behavior under high-pressure conditions (e.g., the $P\bar{6}2c$, $Pnma$, $P2_1/a$, $Pmmn$ and $Pm\bar{3}m$ crystal structures are known). It turns out that correlation effects play an important role in the formation of the electronic and structural properties of FeS under pressure. See, e.g., 
A. V. Ushakov \emph{et al.}, Phys. Rev. B {\bf 95}, 205116 (2017) and
A. O. Shorikov \emph{et al.} Phys. Rev. B {\bf 98}, 094112 (2018).


\bibitem{FeS_sc}
X. F. Lai, H. Zhang, Y. Q. Wang, X. Wang, X. Zhang,
J. H. Lin, and F. Q. Huang,
J. Am. Chem. Soc. {\bf 137}, 10148 (2015);
%
U. Pachmayr, N. Fehn, and D. Johrendt,
Chem. Commun. {\bf 52}, 194 (2016).


\bibitem{FeS_sc_suppress}
X. Lai, Y. Liu, X. L\"u, S. Zhang, K. Bu, C. Jin, H. Zhang,
J. Lin, and F. Huang, 
Sci. Rep. {\bf 6}, 31077 (2016).


\bibitem{FeS_two_domes}
J. Zhang, F.-L. Liu, T.-P. Ying, N.-N. Li, Y. Xu, L.-P. He,
X.-C. Hong, Y.-J. Yu, M.-X. Wang, J. Shen, W.-G. Yang,
and S.-Y. Li, 
npj Quantum Mater. {\bf 2}, 49 (2017).


\bibitem{FeS_Lifshitz}
M. Shimizu, N. Takemori, D. Guterding, and H. O. Jeschke
Phys. Rev. Lett. {\bf 121}, 137001 (2018).


\bibitem{FeCh_DFT}
A. Subedi, L. Zhang, D. J. Singh, and M. H. Du, 
Phys. Rev. B {\bf 78}, 134514 (2008).


\bibitem{FeS_ARPES}
J. Miao, X. H. Niu, D. F. Xu, Q. Yao, Q. Y. Chen, 
T. P. Ying, S. Y. Li, Y. F. Fang, J. C. Zhang, S. Ideta, 
K. Tanaka, B. P. Xie, D. L. Feng, and Fei Chen,
Phys. Rev. B {\bf 95}, 205127 (2017).


\bibitem{FeS_dftdmft}
C. Tresca, G. Giovannetti, M. Capone, and G. Profeta
Phys. Rev. B {\bf 95}, 205117 (2017).


\bibitem{FeS_dft_phonon}
A. Baum, A. Milosavljevi\' c, N. Lazarevi\' c,
M. M. Radonji\' c, B. Nikoli\' c, M. Mitschek,
Z. I. Maranloo, M. \v S\' cepanovi\' c,
M. Gruji\' c-Broj\v cin, N. Stojilovi\' c,
M. Opel, A. Wang, C. Petrovi\' c, Z. V. Popovi\' c,
and R. Hackl,
Phys. Rev. B {\bf 97}, 054306 (2018).


\bibitem{dftdmft_nsc}
W. Metzner and D. Vollhardt, Phys. Rev. Lett. {\bf 62}, 324 (1989);
A. Georges, G. Kotliar, W. Krauth, and M. J. Rozenberg, Rev. Mod. Phys. {\bf 68}, 13 (1996);
%
V. I. Anisimov, A. I. Poteryaev, M. A. Korotin, A. O. Anokhin, and G. Kotliar,
J. Phys. Condens. Matter {\bf 9}, 7359 (1997);
%
G. Kotliar, S. Y. Savrasov, K. Haule, V. S. Oudovenko, O. Parcollet, and C. A. Marianetti,
Rev. Mod. Phys. {\bf 78}, 865 (2006);
%
A. I. Lichtenstein and M. I. Katsnelson, Phys. Rev. B {\bf 57}, 6884 (1998).
%
J. Kune\v s, I. Leonov, P. Augustinsk\' y, V. K\v r\' apek, M. Kollar, and D. Vollhardt,
Eur. Phys. J. Spec. Top. {\bf 226}, 2641 (2017).


\bibitem{dftdmft_sc}
I. Leonov, V. I. Anisimov, and D. Vollhardt,
Phys. Rev. B {\bf 91}, 195115 (2015);
%
I. Leonov,
Phys. Rev. B \textbf{92}, 085142 (2015);
%
I. Leonov, L. Pourovskii, A. Georges, and I. A. Abrikosov,
Phys. Rev. B \textbf{94}, 155135 (2016).


\bibitem{U_in_superconductors}
K. Haule, J. H. Shim, and G. Kotliar,
Phys. Rev. Lett. {\bf 100}, 226402 (2008);
%
M. Aichhorn, L. Pourovskii, V. Vildosola, M. Ferrero,
O. Parcollet, T. Miyake, A. Georges, and S. Biermann,
Phys. Rev. B {\bf 80}, 085101 (2009);
%
S. L. Skornyakov, A. A. Katanin, and V. I. Anisimov,
Phys. Rev. Lett. {\bf 106}, 047007 (2011);
%
S. L. Skornyakov, V. I. Anisimov, and D. Vollhardt, 
Phys. Rev. B {\bf 86}, 125124 (2012);
%
Z. P. Yin, K. Haule, and G. Kotliar,
Nat. Mater. {\bf 10}, 932 (2011);
%
Z. P. Yin, K. Haule, and G. Kotliar,
Nat. Phys. {\bf 7}, 294 (2011);
%
M. Aichhorn, L. Pourovskii, and A. Georges,
Phys. Rev. B {\bf 84}, 054529 (2011);
%
J. M. Tomczak, M. van Schilfgaarde, and G. Kotliar,
Phys. Rev. Lett. {\bf 109}, 237010 (2012);
%
Z. P. Yin, K. Haule, and G. Kotliar,
Phys. Rev. B {\bf 86}, 195141 (2012);
%
P. Werner, M. Casula, T. Miyake, F. Aryasetiawan, A. J. Millis,
and S. Biermann,
Nat. Phys. {\bf 8}, 331 (2012);
%
A. Georges, L. de' Medici, and J. Mravlje,
Annu. Rev. Condens. Matter Phys. {\bf 4}, 137 (2013);
%
M. Hirayama, T. Miyake, and M. Imada,
Phys. Rev. B {\bf 87}, 195144 (2013);
%
C. Zhang, L. W. Harriger, Z. Yin, W. Lv, M. Wang, G. Tan, Y. Song,
D. L. Abernathy, W. Tian, T. Egami, K. Haule, G. Kotliar, and P. Dai,
Phys. Rev. Lett. {\bf 112}, 217202 (2014);
%
A. van Roekeghem, L. Vaugier, H. Jiang, and S. Biermann,
Phys. Rev. B \textbf{94}, 125147 (2016);


\bibitem{Skornyakov_2009}
S. L. Skornyakov, A. V. Efremov, N. A. Skorikov, M. A. Korotin, 
Yu. A. Izyumov, V. I. Anisimov, A. V. Kozhevnikov, and D. Vollhardt, 
Phys. Rev. B {\bf 80}, 092501 (2009).


\bibitem{FeSeHubbard2}
M. Aichhorn, S. Biermann, T. Miyake, A. Georges, and M. Imada,
Phys. Rev. B {\bf 82}, 064504 (2010).


\bibitem{FeSe_compressibility}
P. V. Arribi and L. de' Medici,
Phys. Rev. Lett. {\bf 121}, 197001 (2018).


\bibitem{FeSeHubbard1}
M. D. Watson, S. Backes, A. A. Haghighirad, M. Hoesch, 
T. K. Kim, A. I. Coldea, and R. Valent\'i, 
Phys. Rev. B {\bf 95}, 081106(R) (2017).


\bibitem{Birch}
F. Birch,
Phys. Rev. {\bf 71}, 809 (1947).


\bibitem{GGA}
S. Baroni, S. de Gironcoli, A. Dal Corso, and P. Giannozzi,
Rev. Mod. Phys. {\bf 73}, 515 (2001);
P. Giannozzi, S. Baroni, N. Bonini, M. Calandra, R. Car {\it et al.},
J. Phys.:Condens. Matter {\bf 21}, 395502 (2009).


\bibitem{WannierH}
N. Marzari, A. A. Mostofi, J. R. Yates, I. Souza, and D. Vanderbilt, Rev. Mod. Phys. \textbf{84}, 1419 (2012);
%
V. I. Anisimov, D. E. Kondakov, A. V. Kozhevnikov, I. A. Nekrasov,
Z. V. Pchelkina, J. W. Allen, S.-K. Mo, H.-D. Kim, P. Metcalf, S. Suga,
A. Sekiyama, G. Keller, I. Leonov, X. Ren, and D. Vollhardt,
Phys. Rev. B {\bf 71}, 125119 (2005);
%
Dm. Korotin, A. V. Kozhevnikov, S. L. Skornyakov, I. Leonov, N. Binggeli,
V. I. Anisimov, and G. Trimarchi,
Eur. Phys. J. B {\bf 65}, 91-98 (2008);
%
G. Trimarchi, I. Leonov, N. Binggeli, Dm. Korotin, and V. I. Anisimov,
J. Phys.: Condens. Matter {\bf 20}, 135227 (2008).


\bibitem{ctqmc}
P. Werner, A. Comanac, L. de Medici, M. Troyer, and A. J. Millis,
Phys. Rev. Lett. {\bf 97}, 076405 (2006);
E. Gull, A. J. Millis, A. I. Lichtenstein, A. N. Rubtsov,
M. Troyer, and P. Werner,
Rev. Mod. Phys. {\bf 83}, 349 (2011).


\bibitem{Pade}
H. J. Vidberg and J. W. Serene,
J. Low Temp. Phys. {\bf 29}, 179 (1977).


\bibitem{FeS_bulkmod}
L. Ehm, F. M. Michel, S. M. Antao, C. D. Martin, P. L. Lee, 
S. D. Shastri, P. J. Chupasc, and J. B. Parise,
J. Appl. Cryst. {\bf 42}, 15 (2009).


\bibitem{FeSe_B0}
J. N. Millican, D. Phelan, E. L. Thomas, J. B. Le\~ao, E. Carpenter,
Solid State Commun. {\bf 149}, 707 (2009).


\bibitem{FeSe_structure}
S. Margadonna , Y. Takabayashi, M. T. McDonald, K. Kasperkiewicz,
Y. Mizuguchi, Y. Takano, A. N. Fitch, E. Suard, K. Prassides,
Chem. Commun. (Cambridge) {\bf 43}, 5607 (2008);
%
M. C. Lehman, A. Llobet, K. Horigane, and D. Louca,
J. Phys. Conf. Ser. {\bf 251}, 012009 (2010).


\bibitem{note}
Strictly speaking, our results show that the isothermal 
compressibility $\kappa_T$ calculated within DFT+DMFT as 
$\kappa_T=-1/T (\partial{V}/\partial{P})_T$ diverges at 
the phase transition. This is due to the fact that we 
neglect multiple intermediate-phase transitions when fit 
the total-energy result to the third-order Birch-Murnaghan 
equation of states separately for the low- and high-volume 
regions.

\bibitem{FeSe_magnetic_DFT}
S. L. Skornyakov, I. Leonov, V. I. Anisimov,
JETP Lett. {\bf 103}(4), 265 (2016).


\bibitem{Borisenko_VHS}
J. Phys. Chem. Solids {\bf 72}, 562 (2011).


\bibitem{FeSeTe_PES}
T. Yokoya, R. Yoshida, Y. Utsumi, K. Tsubota, H. Okazaki, 
T. Wakita, Y. Mizuguchi, Y. Takano, T. Muro, and Y. Kato, 
Sci. Technol. Adv. Mater. {\bf 13}, 054403 (2012).

\end{thebibliography}
\end{document}